# Steady shear magnetorheology in $Co_{0.9}Ni_{0.1}$ nanocluster-based MR fluids at elevated temperatures

Injamamul Arief[†], P.K. Mukhopadhyay

*LCMP, Department of Condensed Matter Physics and Material Sciences, S. N. Bose National Centre for Basic Sciences, Salt Lake, Kolkata 700 098, WB, India.*

[†]Email: md.injamamularief@bose.res.in

**Abstract:**

In this paper, we present the study of magnetorheological properties of magnetic fluids containing $Co_{0.9}Ni_{0.1}$ nanocluster that have been measured as a function of both magnetic field and temperature. Co-rich nanoclusters were synthesized by conventional homogeneous nucleation in liquid polyol. Morphological characterization using FESEM revealed the non-aggregated nature of nanoclusters with an average diameter of 450 nm. Crystal structure and room temperature magnetization measurements were performed by powder XRD and vibrating sample magnetometry (VSM). Two MR samples of different particle volume fractions were prepared. Temperature-dependent steady shear MR characterizations for both the samples in the range of $25^0C$-$55^0C$ demonstrated systematic decline of MR parameters with increasing temperatures. The temperature-induced thinning of shear stress and viscosity was explained in terms of change in effective volume fraction and magnetic saturation. To analyze the measured variation in MR response with increasing temperature, suitable temperature-sensitive scaling parameters were also constructed. Finally to generalize the trend, rheological master curves were constructed by using time-temperature-field superposition method.

*Keywords:* $Co_{0.9}Ni_{0.1}$ nanocluster, temperature-dependent magnetorheology, steady shear, scaling parameters, rheological master curves.

## 1. Introduction:

Magneto-rheological fluids or ferrofluids constitute a smart material whose rheological properties, such as apparent viscosity, stiffness and yield stress can be controlled by temperature as well as external magnetic fields. A typical MR fluid consists of magnetic particles of sub-micron to a few microns dimension in suitably chosen carrier fluid [1]. When exposed to magnetic field, field-induced interparticle interaction leads to complex anisotropic structuration within the fluid. This gives rise to field-dependent yield stress [2]. This parameter characterizes viscoplastic properties of MR fluids and is defined as minimum stress applied to initiate the flow. Previous studies have shown the sub-quadratic dependence of yield stress on particle loading in columnar aggregates. In the last few decades, a surge in magnetorheological research has been reported. While most of the publications concentrate on enhancement of material properties and yield stress, very few of them actually deal with temperature-dependence of MR fluids. MR fluids and elastomers are excellent materials for semi-active devices like tunable dampers, brakes, seat suspensions and clutches [3]. An MR device dissipates energy in the form of heat and therefore, increases the temperature [4]. So, operability of MR fluids at elevated temperature is crucial and has to be thoroughly investigated to predict the effect of temperature on the MR effect and the mechanisms that influence the outcome.

Ferromagnetic particles are normally used in magnetic suspension for their higher magnetic saturation, permeability and monodispersity. In addition to conventional carbonyl iron particles (CIP), ferromagnetic binary alloys can also deliver efficiently for their higher saturation magnetizations, low coercivity and remanence and excellent control of shapes and sizes [5]. Previously, CoNi nanosphere and nanowire-based ferrofluids were reported by Gómez-Ramírez et al. [6]. Cubic and spherical FeCo-based MRFs and CoNi nanoflower-based MR fluids were also investigated by the authors [7, 8]. In this paper, we study the influences of temperature on MR effect of the fluids containing CoNi nanoclusters under steady shear rheology. Cobalt rich $Co_{0.9}Ni_{0.1}$ nanoparticles (~450 nm) were synthesized by standard polyol reduction method. Two different MR samples of different volume fractions were prepared by dispersing CoNi into castor oil. We investigate the mechanism that governs non-equilibrium size distribution of internal chain structures with increasing temperature. We also deduce experimental scaling parameters to correlate magnetic field and temperature-dependent shear properties. Scaling laws are generally introduced to rationalize the experimental observations, the dependence of one physical parameter to multiple variables or factors, therefore, provide a useful insight to the internal mechanism that governs interactions within the magnetorheological fluids [9].

## 2. Experimental method and characterization:

The synthesis of spherical $Co_{0.9}Ni_{0.1}$ nanoclusters is carried out by typical one-pot polyol reduction method similar to that described in earlier publications [10]. Briefly, precursor solution is prepared by dissolving Ni (II) and cobalt (II) acetate in ethylene glycol with appropriate molar ratio. Ethylene glycol is acted as both reducing agent and solvent. The polyol reduction is usually carried out at the boiling temperature of ethylene glycol in presence of excess [OH$^-$] and for 2 hr. Afterwards, nanoclusters are isolated and subsequently purified and dried. The magnetorheological (MR) fluids are made by dispersing nanopowder (density 4.8 g/cc) into castor oil (viscosity 0.879 Pa.s at 25$^0$C). Two MR suspensions (MR1 and MR2) are prepared, with dispersed phase concentration of 20 vol% and 15 vol%, respectively. Particle morphology, size and shape are investigated by a field emission scanning electron microscopy (FESEM, Quanta FEG®, FEI) with an energy dispersive x-ray (EDAX) attachment. Crystal structures and phases of the powder samples at room temperature are identified by powder x-ray diffraction using PANalytical X'Pert PRO® diffractometer using monochromatic Cu-K$_\alpha$ radiation ($\lambda$=0.51418 nm).

Magnetorheological measurements with both on- and off-field are investigated using a commercial rheometer (Anton Paar MCR Physica 501®) with magnetorheological attachment (MRD 170®) in strain-controlled mode. The Parallel plate system with plate diameter of 20 mm is used for all measurements. A fixed plate-gap of 1 mm is maintained throughout the measurements. The magnetic field is generated vertically with respect to the direction of flow. For our measurements, magnetic flux density is varied from 0 to 0.6 T. For temperature-dependent MR studies, MCR 501® is attached to a JULABO F25 temperature controlling unit and various temperatures are maintained to high accuracy. Temperature-dependent magnetorheological studies under steady shear (controlled shear rate, CSR) mode are performed at 35$^0$C, 45$^0$C and 55$^0$C, in addition to room temperature. Before any measurement, the sample is pre-sheared at 20 s$^{-1}$ for about 30 s.

## 3. Results and discussions:

### 3.1. Morphological characterizations:

The surface morphology and average particle diameter are studied using FESEM image of as synthesized Co$_{0.9}$Ni$_{0.1}$ powder. Nanoclusters are non-aggregated and nearly-spherical in size with an average diameter of 450 nm. The unique surface morphology of nanoclusters can be understood in terms of nanoparticle growth in polyol reduction. Evolution of nanoclusters occurs through distinct homogeneous nucleation followed by growth phases [10]. The EDX spectral studies are performed on powdered nanoclusters and shown in Fig. 1C. It confirmed the composition of nanocluster with high Co-content which is similar to the initial molar ratio of metal salts, i.e. [Co]: [Ni] = 9:1. Spot EDX of nanoclusters further established compositional homogeneity. The crystal structures and phases of the alloy nanoclusters are determined by powder x-ray diffraction as shown in Fig. 1B. The featured peaks at θ= 44.6$^0$, 51.7$^0$, 77.6$^0$ are assigned to fcc crystal structure while significant presence of hcp Co phases at θ =42$^0$, 44.8$^0$ and 47.5$^0$ are also indexed. Co-rich CoNi alloys crystallize in mixed fcc and hcp phases [10-12]. Room temperature magnetometric study of sample pellets (pressed powder) indicates ferromagnetic character of the sample at room temperature (Fig. 1D). The saturation magnetization (M$_s$), remnant magnetization (M$_r$) and coercive field (H$_c$) values are calculated to be 119.5 emu/g, 21 emu/g and 142 Oe, respectively.

### 3.2. Temperature-dependent magnetorheology and scaling parameters:

The effect of temperature on the apparent viscosities of MR fluids is investigated by effective apparent viscosity ratio, η$_{eff}$ and is defined as the ratio of change in apparent viscosity with increasing temperature to that of viscosity measured at a reference temperature:

$$\eta_{eff} = \frac{\eta_{app}(T_{25C},\dot{\gamma}) - \eta_{app}(T_{55C},\dot{\gamma})}{\eta_{app}(T_{25C},\dot{\gamma})} \quad (1)$$

Where $\eta_{app}(T_{55C}, \dot{\gamma})$ and $\eta_{app}(T_{25C}, \dot{\gamma})$ are the apparent viscosity at $55^0$C and at room temperature, respectively. A numerically higher $\eta_{eff}$ signifies higher temperature sensitivity and lower magnitude implies the inferior response with respect to temperature [13]. Figure 2 shows effective viscosity ratio as a function of constant shear rate for both samples MR1 and MR2 under different magnetic field, in addition to zero field. It can be observed that the zero-field value of $\eta_{eff}$ for MR1 and MR2 are much higher than that of the field-dependent values. Also $\eta_{eff}$ of MR2 is slightly higher in magnitude throughout the range of shear rate applied, compared to MR1. The initial and final values of $\eta_{eff}$ for MR1 are 0.76 and 0.79 and for MR2 are 0.83 and 0.81, respectively. With an increase in particle concentration of about 33% by volume, the effect of temperature to the apparent viscosity is lessened somewhat. Although the change is not very high, it can be explained in terms of induced microstructures formed even in the absence of field. Due to higher particle concentration, matrix fluid is entrapped in between the induced aggregates formed at rest. Since suspension viscosity is a strong function of carrier fluid viscosity in absence of magnetic field, relative change in fluid viscosity with temperature is lower due to lower effective concentration of free carrier fluid. With increasing field, a similar lowering of $\eta_{eff}$ is also observed for both suspensions. At sufficiently higher shear rate (>200 s$^{-1}$), field induced structures tend to collapse and suspension starts flowing. Beyond this shear rate, $\eta_{eff}$ for MR1 and MR2 under different magnetic field dominated by the matrix fluid viscosity and become identical as fluids flow indefinitely due to absence of any microstructure aggregates.

The effect of temperature on shear stress of MR fluids (MR1 and MR2) is presented in Fig. 3 for different magnetic fields. The measurements were performed within the temperature range of $25^0$C-$55^0$C. Shear stress (as a function of shear rate) is observed to decrease systematically with elevation of temperature and is applicable to all the values of magnetic fields. The dynamic yield stress ($\tau_{yd}$) is calculated by fitting the shear stress vs. shear rate curves with Bingham equation, $\tau = \tau_b + \eta_p \dot{\gamma}$, where $\eta_p$ is plastic viscosity [14]. Therefore, this is the extrapolated value at which the plastic flow would have occurred first. Since this is the major portion in the stress-shear rate curve in which the MR devices operate, it is a very important parameter. The effect of temperature on magnetorheological yield stress is evaluated for five levels of magnetic fields and shown in Fig. 4A and 4B for MR1 and MR2, respectively. It is obvious that yield stress depends significantly on temperature. Therefore, an extended form of Bingham equation can be rewritten as follows:

$$\tau(\dot{\gamma} B, T) = \tau_{yd}(B,T) + \eta_p \dot{\gamma} \quad (2)$$

The trend in dynamic yield stress with temperature is explained by an exponential decay function for $\tau_{yd}$ and the fits are reasonably good with the fitting function $\tau_{yd} = Ke^{-bT}$ where K and b are empirical constant, T is temperature in K. The exponential relationship between yield stress and temperature can be originated from the Arrhenius type relationship between viscosity and temperature: $\eta = Ae^{-\frac{E_a}{RT}}$ where, R is universal gas constant, A is a constant, T is temperature in K and $E_a$ is activation energy to initiate flow. This relation provides reasonable justification to the observed trend in viscosity with temperature and the fact that viscosity and yield stress are associated with the flow behavior of the material, a similar exponential relationship can also be used for dynamic yield stress as shown in Fig. 4A and 4B [13]. The proposed model provides a good agreement with all experimental data for both MR1 and MR2. The change in yield stress with temperature is more pronounced for MR2 with the value of exponential decay constant (*b*) of 0.041 whereas, a lesser rapid change in $\tau_{yd}$ is observed for MR1 with the *b* value of 0.005. One possible explanation for a more rapid reduction in yield stress may be the effect of temperature on the volume fraction of ferromagnetic CoNi nanospheres dispersed in carrier fluid. Since castor oil has higher thermal expansion coefficient than dispersed phase, with increasing temperature the effective volume fraction occupied by particles decreases [15]. Using the following relation, one can calculate normalized thermal sensitivity of suspensions:

$$S_\phi = \lim_{\Delta T \to 0} \frac{1}{\phi_0} \frac{\phi_f - \phi_0}{\Delta T} = (1-\phi_0)(\alpha_p - \alpha_f) \quad (3)$$

Where $\phi_0$, $\phi_f$, $\alpha_p$ and $\alpha_f$ are initial volume fraction, effective volume fraction, volumetric thermal expansion coefficients for particles and carrier fluid, respectively. From the relation, it is obvious that MR2 shows higher normalized thermal sensitivity. Below a critical concentration of φ= 0.3, a linear variation in yield stress with volume fraction is observed. Therefore, MR2 with higher thermal sensitivity is expected to demonstrate a stronger decay in temperature-dependent yield stress.

The trend in dynamic yield stress at a particular temperature is also illustrated as a function of magnetic flux (Fig. 5). In both cases, the trend in viscoelastic parameters with temperature is worth noted. The observations indicate a temperature-thinning effect. This can be qualitatively explained by thermal vibration of the suspended nanoclusters with increasing temperature. According to Li et al. [4], magnetic particles acquire thermal energy at a given temperature. With increasing magnetic field, alignment of particles into chain-like structures disrupts the randomness. With increasing temperatures, due to increase in thermal energy, structures are appeared to be broken. Therefore, a reduction in magnetorheological parameters is followed. This explanation, however fails to justify very large value of parameter λ. It is a dimensionless quantity and defined as a ratio of magnetostatic force to characteristic Brownian force acting on a particle kT/a:

$\lambda = \dfrac{\pi \mu_0 a^3 M^2}{6kT}$ Where a, M, T are radius, saturation magnetization and absolute temperature; $\mu_0$ and k are vacuum permeability and Boltzmann constant, respectively. Typical λ value for the particle size of 450 nm at $55^0$C is ~O ($10^7$) which implies a very negligible Brownian force contribution to stress. Therefore, thermal vibration of particle chains is insufficient to cause the observe changes in MR parameters. Hence, to understand the effect of temperatures, quantitative studies of the changes in rheological parameters are inevitable. The correlation is a direct output of a number of parameters describing the trend in yield stress, magnetization and matrix viscosity as a function of temperature. Therefore, it is necessary to define those parameters and establish a quantitative analogy for the trends described above.

The reduction in yield stress ($\tau_{yd}$) with temperature is described by introducing an average normal yield stress sensitivity parameter $\langle S_\tau \rangle$ as following:

$$\langle S_\tau \rangle = \left( \dfrac{1}{\tau_{yd}^0} \right) \dfrac{\Delta \tau_{yd}}{\Delta T} \qquad (4)$$

Where $\tau_{yd}^0$ is yield stress at a reference temperature (25$^0$C) and reference magnetic field (0.125 T) and $\Delta\tau_{yd}$ is the change in yield stress correspond to change in temperature ($\Delta T$). The experimentally calculated magnitude of <$S_\tau$> under different temperatures and magnetic flux are listed in Table 1. The determined values for <$S_\tau$> are observed to be comparable to other experimental results performed previously by Li et al., and Ocalan et al. [4, 15] Another important temperature-dependent parameter is magnetization. Magnetic force is a function of temperature and arises due to thermal within particles. For ferromagnetic particles, saturation magnetization is a monotonically decreasing function of temperature below Curie point. For $Co_{0.9}Ni_{0.1}$, thermal sensitivity is expected to be very small as operating temperature window during magnetorheological studies are far below Curie temperature. Using the relation by Crangle and Goodman, the average sensitivity in saturation magnetization for the temperature range 25$^0$C-55$^0$C is calculated to be:

$$S_M = \frac{1}{M_S^0}\frac{\partial M_S}{\partial T} = 1.4\times10^{-4}/^0C. \quad (5)$$

The off-state viscosity of MR fluid is also strongly dependent on temperature and is related to carrier fluid viscosity. In the present study, castor oil was used for the preparation of MR suspensions. The change in carrier fluid viscosity with temperature is described in Fig. 6. It is observed that the trend follows Arrhenius relationship. The suspension viscosity of MR fluid in absence of magnetic field can also be approximated by Arrhenius equation:

$$\eta_\infty(T) = \eta_\infty(T_0)e^{\frac{\Delta H}{R}\left(\frac{1}{T}-\frac{1}{T_0}\right)} \quad (6)$$

Where $\eta_\infty(T_0)$ and $\eta_\infty(T)$ define infinite shear viscosity at a reference temperature $T_0$ and at any arbitrary temperature $T$, $\Delta H$ is activation energy to flow and R is universal gas constant. To determine limiting viscosities at infinite shear rates, the experimental data of viscosity as a function of shear rate in absence of field for MR1 and MR2 were fitted with Cross model equation (Fig. 7A and 7B), which describes pseudoplastic flow with asymptotic viscosities at zero ($\eta_0$) and infinite ($\eta_\infty$) shear rates:

$$\eta(\dot{\gamma}) = \eta_\infty + (\eta_0 - \eta_\infty) \frac{1}{1+(\lambda \dot{\gamma})^m} \quad (7)$$

Where, $\eta(\dot{\gamma})$, $\lambda$ and m are apparent viscosity, characteristic time constant with unit of time and dimensionless exponent of Cross equation signifying the width of transition between zero-shear viscosity and power-law plateau region of the viscosity curve, respectively [16]. The quantities $\eta_0$, $\eta_\infty$, $\lambda$ and m are all fitting parameters and shown in table 2.

Therefore, all the parameters sensitive to temperature and magnetic field collectively account for the significant changes in magnetorheological parameters at high temperatures. In order to generalize the collective response of the fluid systems at high temperature and magnetic field, the data from Fig. 3 can be made to collapse into a single master curve. The fact that isothermal curves for shear stress variation with different magnetic field as a function shear rate can be superimposed implies the dynamically self-similar system response [9]. The rheogram at the temperature $25^0$C and magnetic field 0.125T was chosen to be the reference curve. All other rheograms are made to collapse on to the reference curve. For this, scaling of shear stress and shear rate is necessary and can be evaluated from the individual shift parameters according to the response of the fluids described above [17]. The yield stress shift parameter ($a_Y$), magnetization shift factor ($a_M$) and thermo-viscous shift parameter ($a_T$) is defined as follows:

1. $a_Y = 1 + \langle S_\tau \rangle \Delta T$ Where $\langle S_\tau \rangle$ is average yield stress sensitivity parameter and $\Delta T$ is change in temperature with respect to reference temperature $T_0$ ($25^0$C).

2. $a_M = \dfrac{M}{M_0}$ Where, $M_0$ is reference magnetization at reference magnetic flux (0.125 T) corresponds to reference yield stress and is obtained from powder magnetization data.

3. Thermo-viscous shift parameter, $a_T = \dfrac{\eta_\infty(T)}{\eta_\infty(T_0)} = e^{\frac{\Delta H}{R}\left(\frac{1}{T} - \frac{1}{T_0}\right)}$

It is defined as the ratio of fluid viscosity at infinite shear rate ($\eta_\infty$) under a certain temperature to that of the $\eta_\infty$ at reference temperature ($T_0$). Since suspension viscosity arises due to the viscous drag on particle chain network, therefore, $a_T$ is defined by the ratio of viscosity of fluids under flowing condition and not by the ratio of zero shear viscosity.

To generate satisfactory time-temperature-field superposition, the above shift factors should be utilized to calculate the reduced or scaled shear stress and shear rate. Therefore, scaled shear stress ($\tau_{sc}$) and shear rate ($\dot{\gamma}_{sc}$) can be represented as follows:

$$\tau_{sc} = \frac{\tau}{a_Y a_M} \quad (8)$$

$$\dot{\gamma}_{sc} = \frac{\dot{\gamma}}{a_Y a_M} a_T \quad (9)$$

Where, $\tau$ and $\dot{\gamma}$ represent actual shear stress and shear rate respectively.

The resulting master curves for MR1 and MR2 are shown in Fig. 8. The superposition is very well for both MR1 and MR2. It can be concluded that physical processes responsible for changes in temperature-dependent MR behavior can be successfully explained by magnetic and thermal shift parameters. The collapse of the rheograms into a single master curve indicates a better correlation of reference yield stress ($\tau_{yd}^0$) to reference magnetization ($M_0$) and reference temperature ($T_0$). Master curves for MR1 and MR2 at reference temperature of $25^0C$ represent material function (shear stress) as they would have been calculated at broad ranges of operating shear rates. The collapsing of data points for samples under various magnetic fields and temperature implies the extension of shear rate range which was beyond the experimental scope. Therefore, the reference yield stress at a reference temperature and magnetic field is known, magnetorheological shear stress at any arbitrary temperature and field can be calculated using the master equation: $\tau(T, M) = \tau_{yd}^0 (1 + \langle S_\tau \rangle \Delta T) \left( \frac{M}{M_0} \right) \quad (10)$

For both MR1 and MR2, concentration of suspension plays an important role. As stated earlier, with increasing temperature, effective volume fraction of particles in suspensions decreases. This change in volume fraction may be the reason for decrease in temperature-dependent yield stress [15]. Using equation (3), it can be shown that the average sensitivity in yield stress is in fact of the same order with the thermal sensitivity in volume fraction. Hence, it can be concluded that the change in yield stress with increasing temperature is directly associated to the change in effective volume fraction of suspension. When particle concentration is lowered by 5 vol%, a stronger correlation is expected due to higher normalized thermal sensitivity in volume fraction for MR2. The experimental results seem to be in good agreement with this claim.

## 4. Conclusion:

In this paper, we have shown the steady shear thermorheological characterizations of two MR suspensions with different $Co_{0.9}Ni_{0.1}$ concentration. The temperature-dependent MR characterization has shown a decline in MR properties at elevated temperatures for both samples. This is expected as at higher temperature, due to large magnitude of volumetric thermal expansion coefficient of carrier fluid, effective particle volume fraction is decreased. Considering the thermal sensitivity of yield stress, magnetic saturation of powder $Co_{0.9}Ni_{0.1}$ and matrix fluid viscosity at elevated temperature, we have introduced suitable scaling parameters. Combining all the shift factors, we have shown that time-temperature-field superposition method is applicable for the two MR suspensions with different $Co_{0.9}Ni_{0.1}$ volume fractions. The master curves generated for both the fluids with respect to a reference temperature ($T_0$), reference magnetization ($M_0$) and reference yield stress ($\tau_{yd}^0$), indicate a better correlation between the individual sensitivity parameters that are affected differently with increasing temperature. A good superposition of data points for samples under various magnetic fields and temperature implies that the systems are dynamically identical. Although the behavior of each system is governed by similar physical mechanism, the process can be extended at different extent for each sample.

**Acknowledgments:**

One of the authors, IA thanks CSIR, India for the award of Senior Research Fellowship (SRF). The authors also thank Dr. Surajit Dhara, Associate Professor, School of Physics, Central University of Hyderabad, India for magnetorheological measurements.

**Figures with captions:**

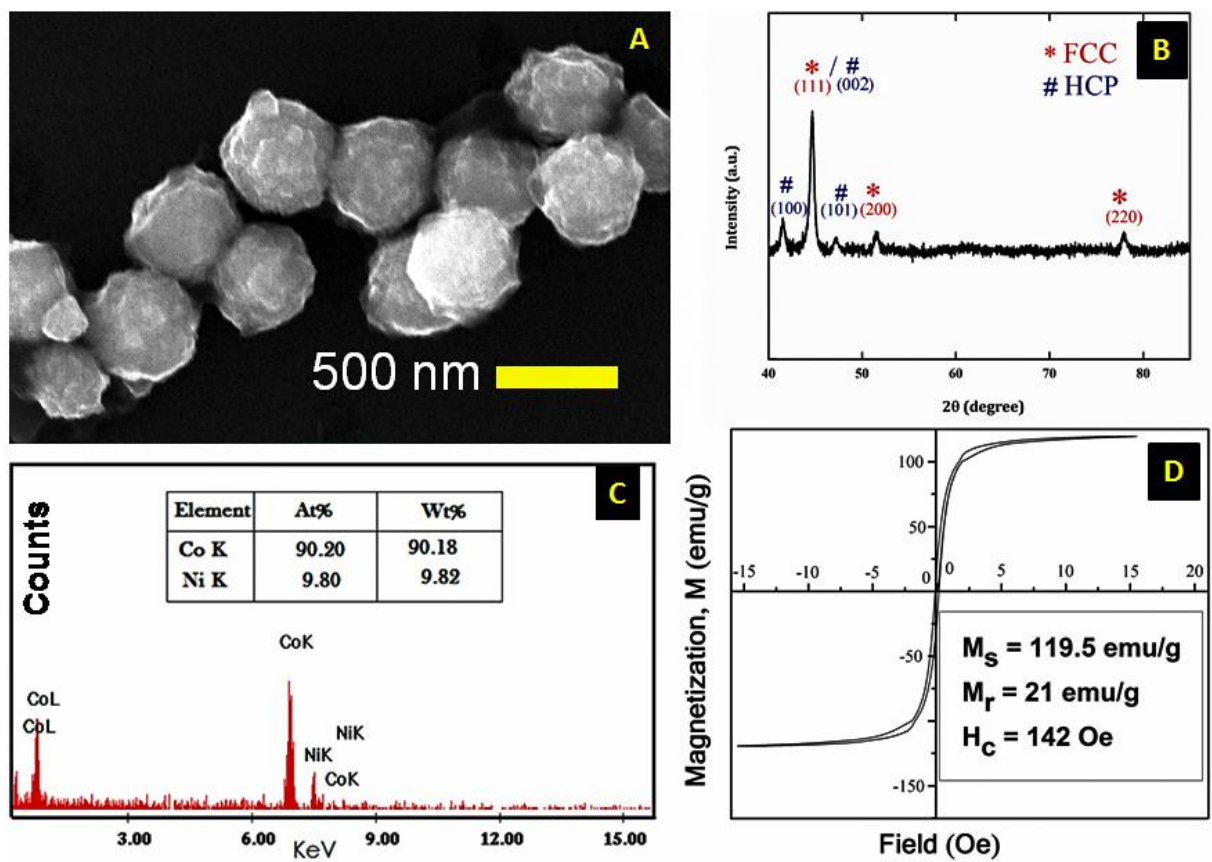

**Fig. 1.** (A) FESEM image of $Co_{0.9}Ni_{0.1}$ nanoclusters, average diameter of nearly spherical nanoclusters was found to be 450 nm; (B) powder X-ray diffraction peaks of as-synthesized $Co_{0.9}Ni_{0.1}$ nanoclusters. All the peaks are assigned and it is clearly observed that structure crystallizes in mixed fcc-hcp phases; (C) EDX spectra of the sample with desired composition and (D) magnetization curve at T=298K for $Co_{0.9}Ni_{0.1}$ nanoclusters.

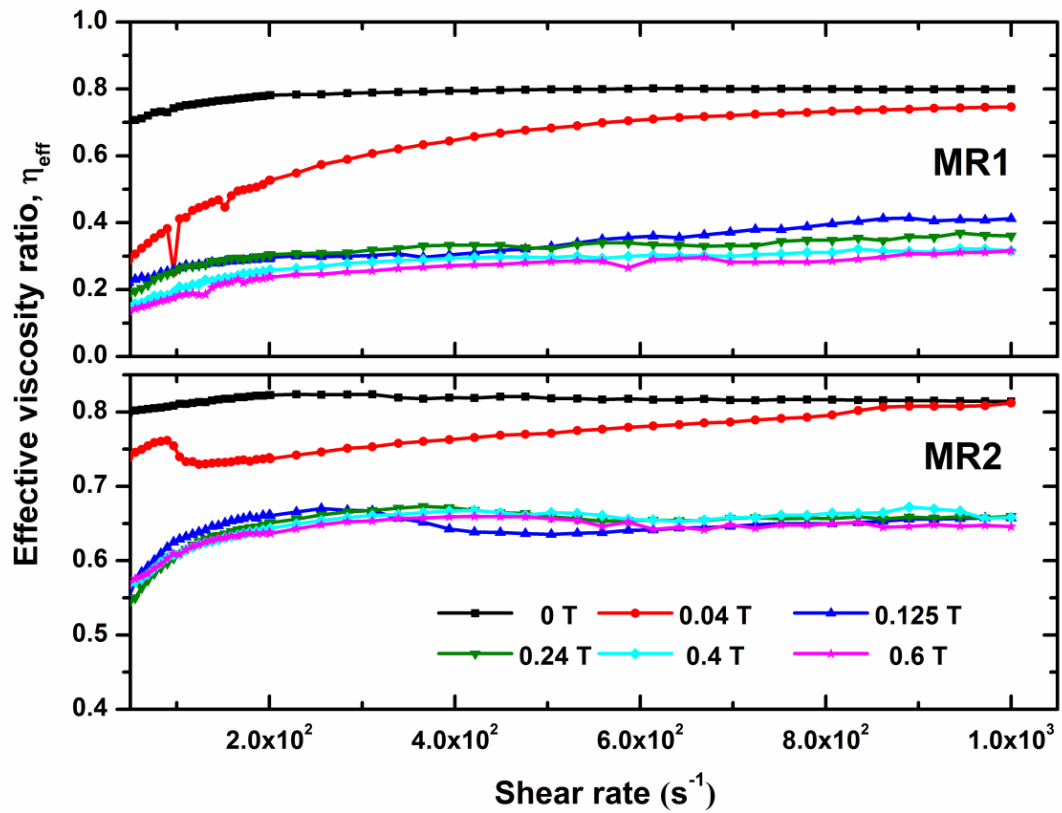

**Fig. 2.** Effective apparent viscosity ratio ($\eta_{eff}$) versus shear rate plots of MR1 and MR2 under various magnetic fields, in addition to zero magnetic fields.

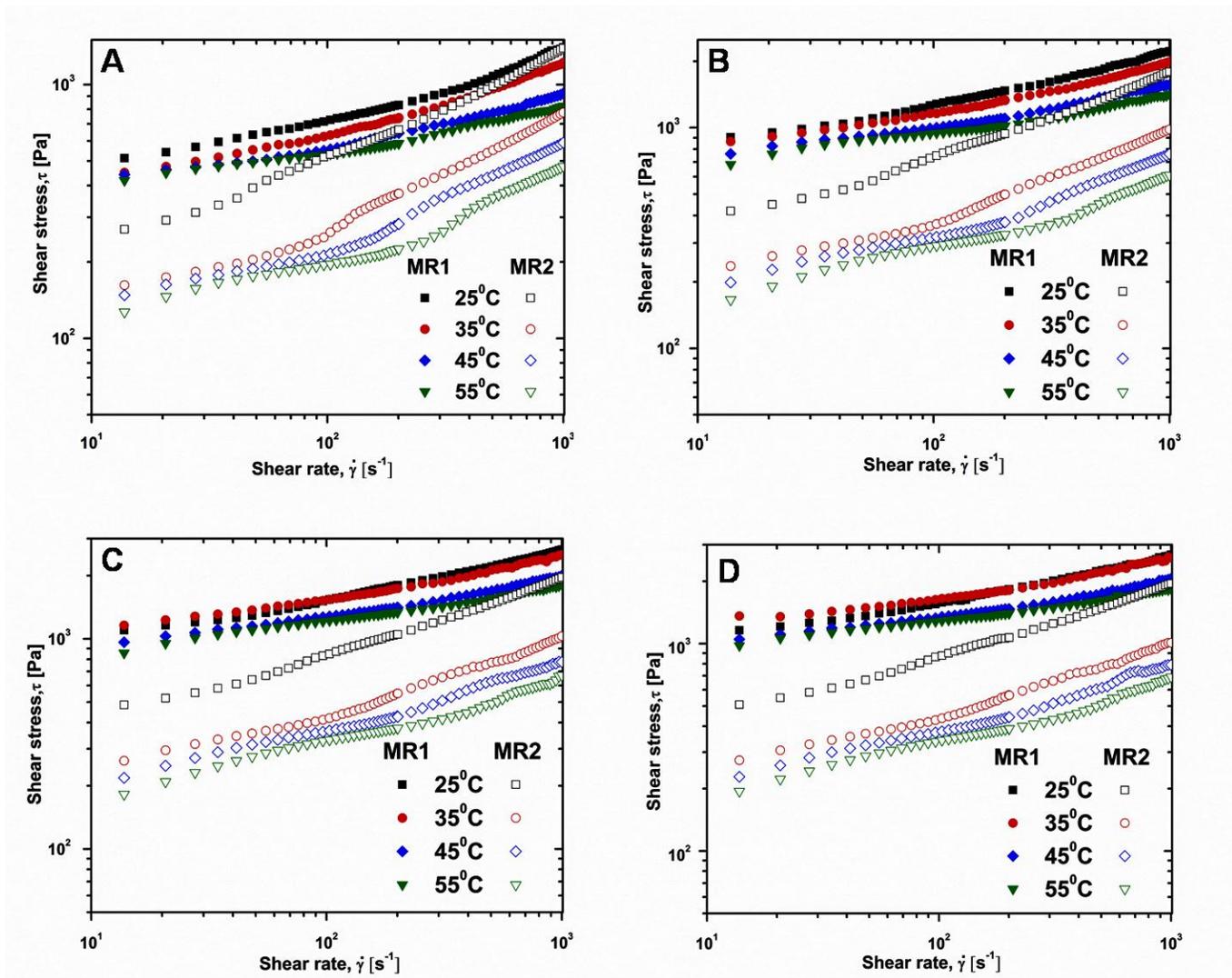

**Fig. 3.** Flow curves of magnetorheological shear stress (τ) as a function of shear rate $(\dot{\gamma})$ and temperature for MR1 (closed symbols) and MR2 (open symbols) under various magnetic fields: (A) 0.125T, (B) 0.24T, (C) 0.4T and (D) 0.6T.

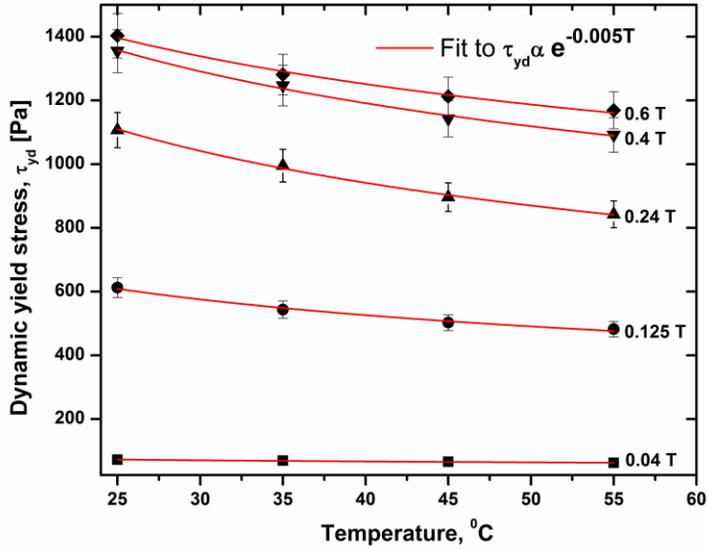

**Fig. 4A.** Bingham or dynamic yield stress plotted as a function of temperature ($25^0$C-$55^0$C) under various magnetic fields for MR1. Data fitting lines represent exponential decay with temperature.

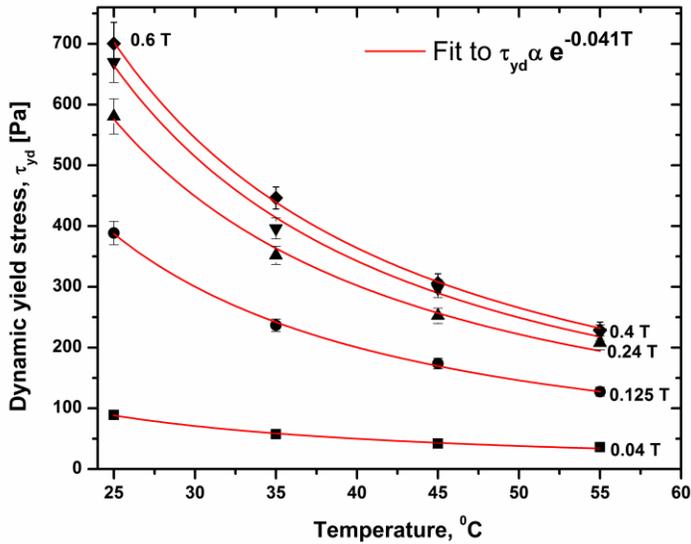

**Fig. 4B.** Dynamic yield stress values of MR2 decay exponentially with respect to increasing temperatures. Fit lines represent a stronger exponential dependence with temperature.

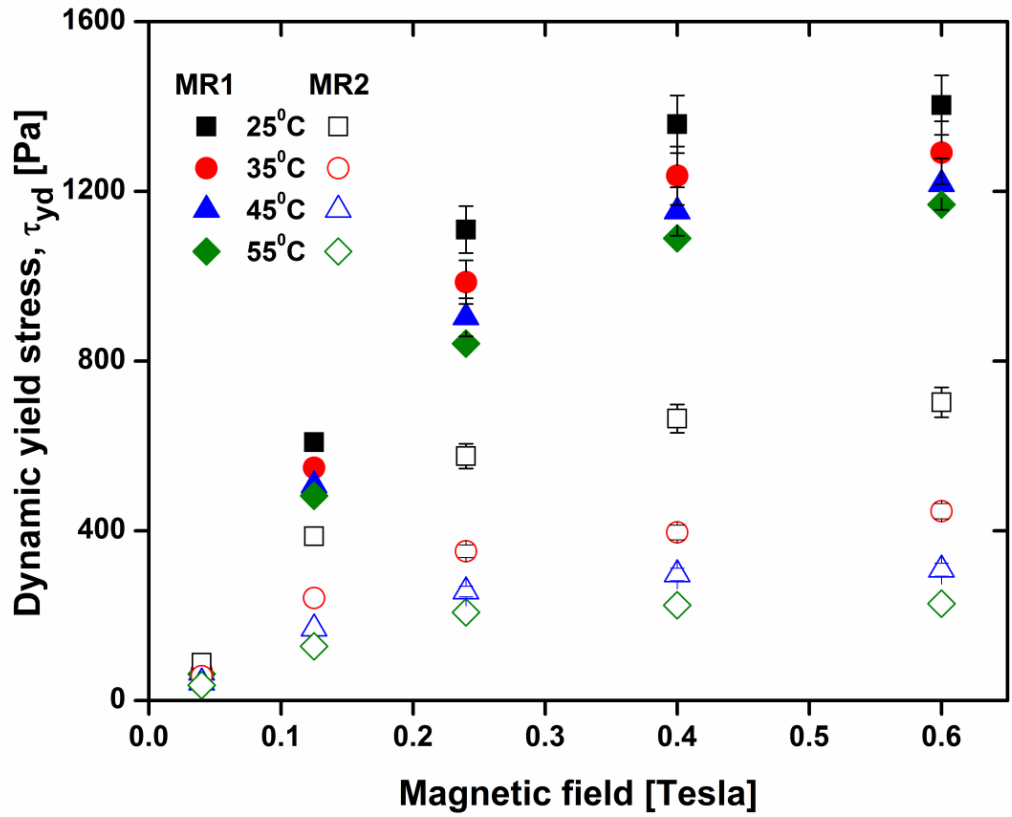

**Fig. 5.** Dynamic yield stresses of MR1 (closed symbols) and MR2 (open symbols) plotted as a function of magnetic field under different temperatures.

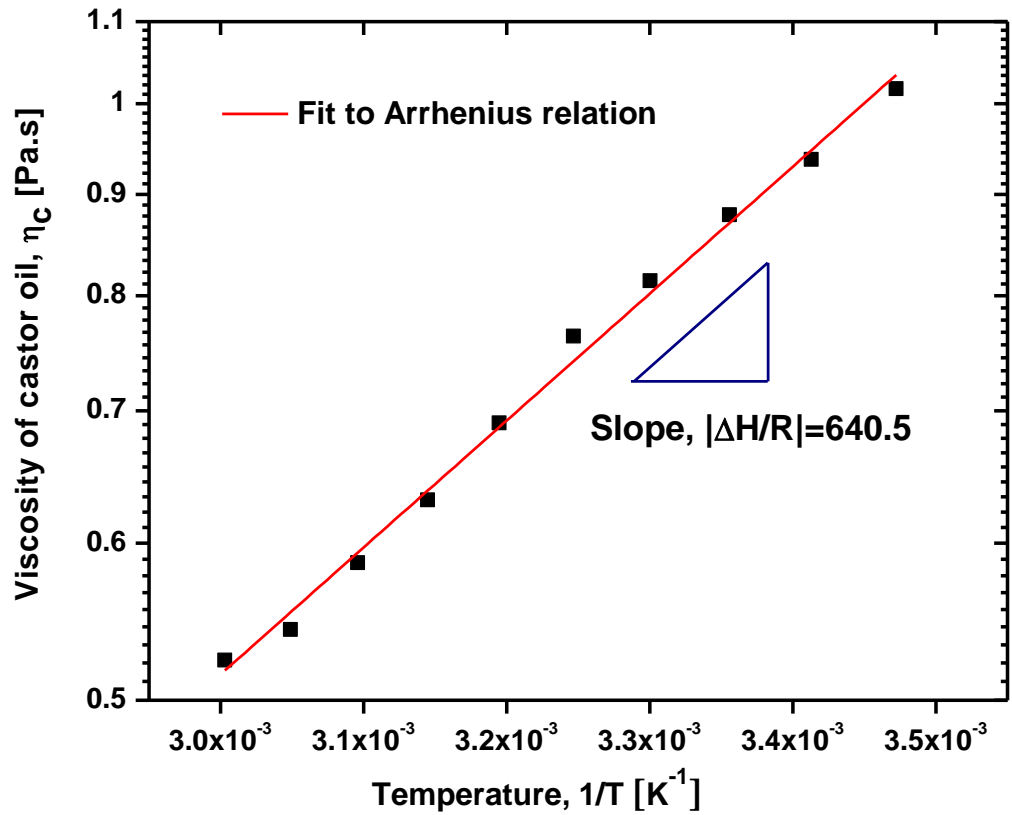

**Fig. 6**. Viscosity of castor oil is plotted as function of inverse absolute temperature (K$^{-1}$). Fitted line represents Arrhenius equation. From the slope of the fitted curve, activation energy for carrier fluid is calculated to be 5.3 kJ/mol.

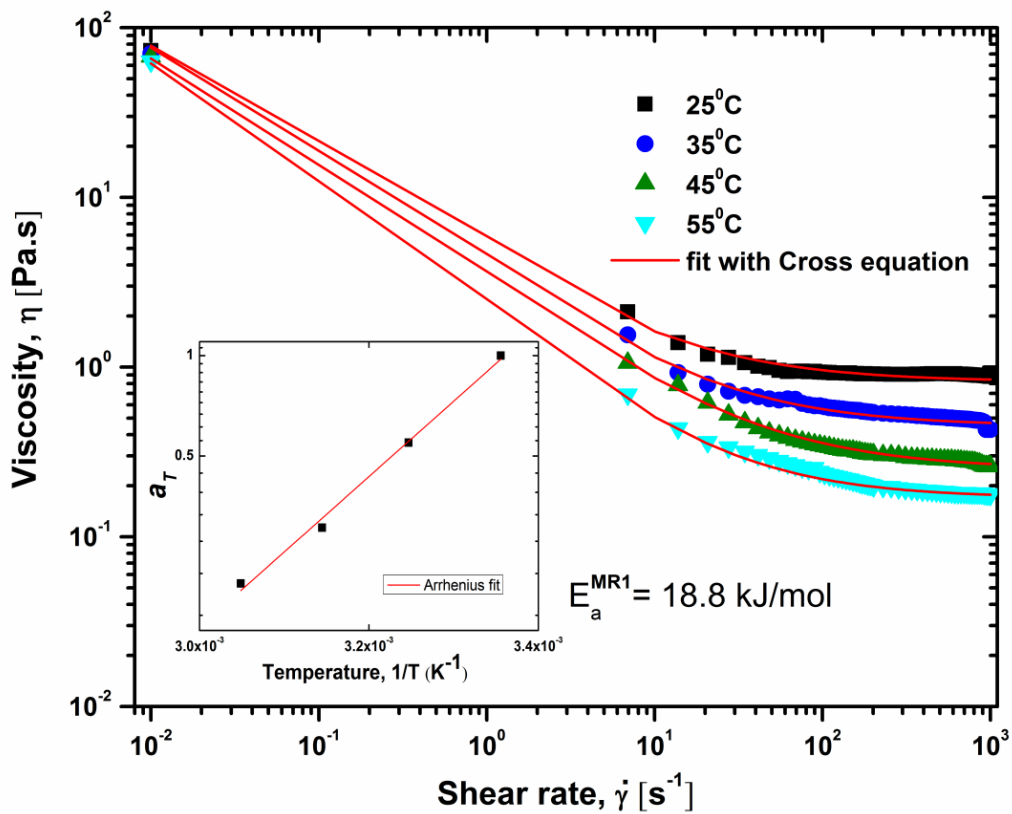

**Fig. 7A.** Zero-field apparent viscosities of MR1 are plotted as a function of shear rate under different operating temperature. Fitted lines represent the same calculated from Cross rheological model equation. In the inset, thermo-viscous shift parameter $a_T$ is plotted against inverse of absolute temperature (K$^{-1}$). Fitted line represents Arrhenius equation.

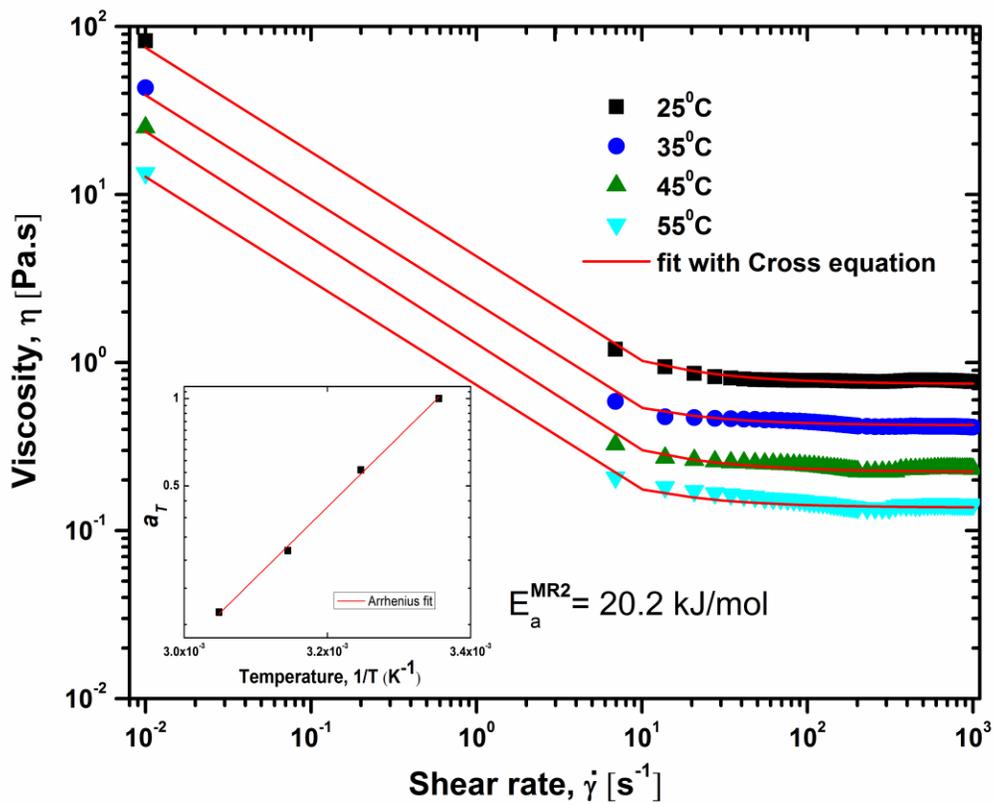

**Fig. 7B**. Off-field apparent viscosities of MR2 versus shear rate are plotted under various operating temperature. Fitted lines show the prediction from Cross rheological model equation; (inset) thermo-viscous shift parameter $a_T$ is plotted against inverse of absolute temperature (K$^{-1}$). Fit line represents Arrhenius equation.

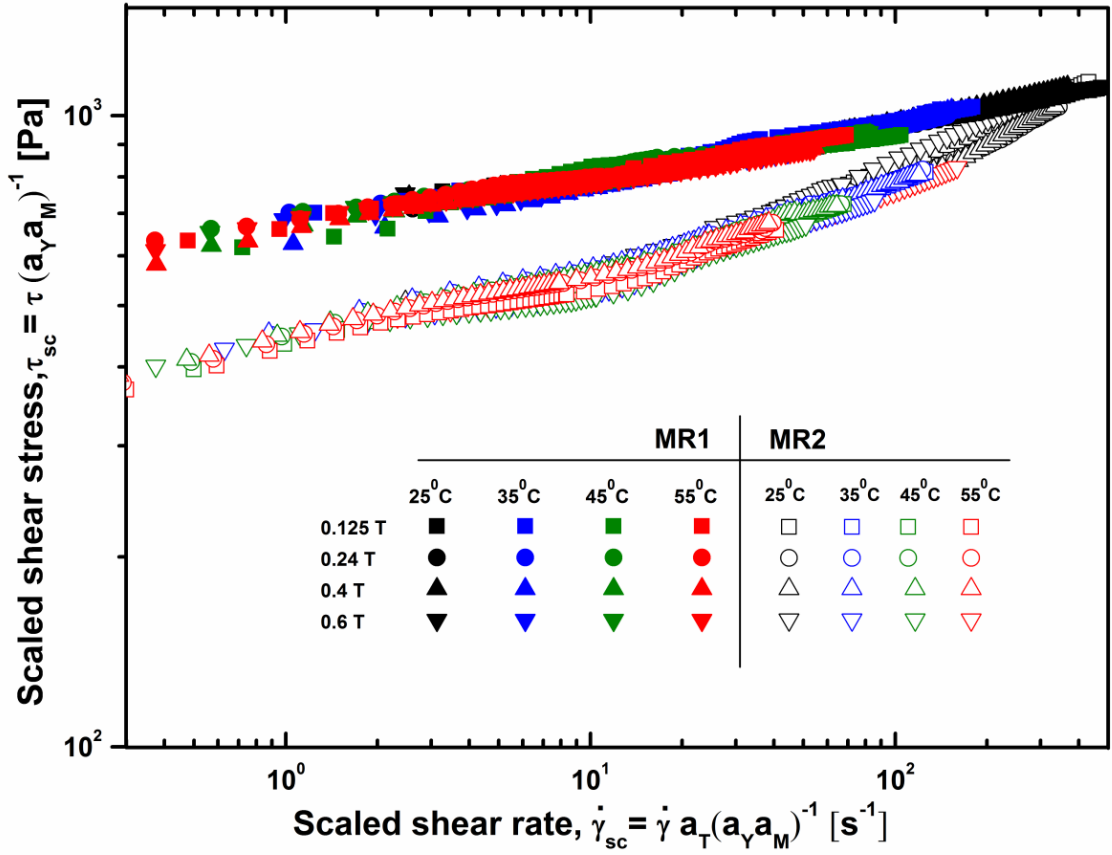

**Fig. 8.** Magnetorheological master curves for MR1 (closed symbols) and MR2 (open symbols) are shown as a function of scaled shear stress ($\tau_{sc}$) and shear rate ($\dot{\gamma}_{sc}$). The scaling parameters were derived considering the dependence of stress parameters on temperature and magnetic field. The $a_Y$, $a_M$ and $a_T$ represent yield stress shifting parameter, magnetic shift and thermo-viscous shift parameters, respectively.

**Table 1** Yield stress sensitivity parameter and scaling factor as a function of flux densities and temperatures.

| | MR1 | | | | | |
|---|---|---|---|---|---|---|
| **B (T)** | Average yield stress sensitivity, $\|<S_\tau>\|/10^3/^0C$ | | | Yield stress scaling factor, $a_Y$ | | |
| | $25^0C$-$35^0C$ | $25^0C$-$45^0C$ | $25^0C$-$55^0C$ | $25^0C$-$35^0C$ | $25^0C$-$45^0C$ | $25^0C$-$55^0C$ |
| 0.125 | 11.27 | 8.97 | 7.08 | 1.113 | 1.179 | 1.212 |
| 0.24 | 10.09 | 9.50 | 7.96 | 1.10 | 1.190 | 1.239 |
| 0.4 | 7.97 | 7.81 | 6.47 | 1.079 | 1.156 | 1.194 |
| 0.6 | 8.70 | 6.79 | 5.56 | 1.087 | 1.135 | 1.166 |
| | MR2 | | | | | |
| **B(T)** | Average yield stress sensitivity, $\|<S_\tau>\|/10^3/^0C$ | | | Yield stress scaling factor, $a_Y$ | | |
| | $25^0C$-$35^0C$ | $25^0C$-$45^0C$ | $25^0C$-$55^0C$ | $25^0C$-$35^0C$ | $25^0C$-$45^0C$ | $25^0C$-$55^0C$ |
| 0.125 | 39.02 | 27.64 | 22.37 | 1.390 | 1.553 | 1.671 |
| 0.24 | 39.38 | 28.26 | 21.39 | 1.394 | 1.565 | 1.641 |
| 0.4 | 40.85 | 27.80 | 22.15 | 1.408 | 1.556 | 1.664 |
| 0.6 | 36.27 | 28.14 | 22.47 | 1.363 | 1.563 | 1.674 |

**Table 2** Cross equation fitting parameters for off-field viscosities of MR1 and MR2 with increasing temperature.

| Temperature | 25$^0$C | | 35$^0$C | | 45$^0$C | | 55$^0$C | |
|---|---|---|---|---|---|---|---|---|
| Fitting parameter | MR1 | MR2 | MR1 | MR2 | MR1 | MR2 | MR1 | MR2 |
| $\eta_0$ (Pa.s) | 156.4 | 137.37 | 150.4 | 94.4 | 148.3 | 65.5 | 135.7 | 35.87 |
| $\eta_\infty$ (Pa.s) | 0.82 | 0.745 | 0.45 | 0.423 | 0.25 | 0.223 | 0.17 | 0.137 |
| $\lambda$ (s) | 33.5 | 58.6 | 35.6 | 98.24 | 41.7 | 107.34 | 57.8 | 115.54 |
| m | 0.76 | 0.83 | 0.78 | 0.81 | 0.78 | 0.8 | 0.76 | 0.82 |